\documentclass[10pt,conference]{IEEEtran}
\IEEEoverridecommandlockouts
\usepackage{cite}
\usepackage{amsmath,amssymb,amsfonts}
\usepackage{algorithmic}
\usepackage{graphicx}
\usepackage{textcomp}
\usepackage[dvipsnames]{xcolor}
\usepackage{hyperref}
\def\BibTeX{{\rm B\kern-.05em{\sc i\kern-.025em b}\kern-.08em
    T\kern-.1667em\lower.7ex\hbox{E}\kern-.125emX}}
\usepackage{listings}
\usepackage{makecell}
\usepackage{float}
\usepackage{multirow}
\usepackage{cleveref}
\usepackage[poster]{tcolorbox}
\usepackage{lscape}
\usepackage{url}
\usepackage[utf8]{inputenc}
\definecolor{applegreen}{rgb}{0.0, 0.5, 0.0}

\newcommand\YAMLcolonstyle{\color{red}\mdseries}
\newcommand\YAMLkeystyle{\color{black}\bfseries}
\newcommand\YAMLvaluestyle{\color{blue}\mdseries}

\makeatletter
\newcommand\language@yaml{yaml}

\expandafter\expandafter\expandafter\lstdefinelanguage
\expandafter{\language@yaml}
{
  keywords={true,false,null,y,n},
  keywordstyle=\color{darkgray}\bfseries,
  basicstyle=\YAMLkeystyle,                                 
  sensitive=false,
  comment=[l]{\#},
  morecomment=[s]{/*}{*/},
  commentstyle=\color{purple}\ttfamily,
  stringstyle=\YAMLvaluestyle\ttfamily,
  moredelim=[l][\color{orange}]{\&},
  moredelim=[l][\color{magenta}]{*},
  moredelim=**[il][\YAMLcolonstyle{:}\YAMLvaluestyle]{:},
  morestring=[b]',
  morestring=[b]",
  literate =    {---}{{\ProcessThreeDashes}}3
                {>}{{\textcolor{red}\textgreater}}1     
                {|}{{\textcolor{red}\textbar}}1 
                {\ -\ }{{\mdseries\ -\ }}3,
}

\lst@AddToHook{EveryLine}{\ifx\lst@language\language@yaml\YAMLkeystyle\fi}
\makeatother

\newcommand\ProcessThreeDashes{\llap{\color{cyan}\mdseries-{-}-}}

\newcommand\Tstrut{\rule{0pt}{2.6ex}}         
\newcommand\Bstrut{\rule[-0.9ex]{0pt}{0pt}}   

\makeatletter
\newcommand\footnoteref[1]{\protected@xdef\@thefnmark{\ref{#1}}\@footnotemark}
\makeatother

\begin{document}

\title{LLMs for Automated Unit Test Generation and Assessment in Java: The \textsc{AgoneTest} Framework}

\author{
\IEEEauthorblockN{
Andrea Lops\IEEEauthorrefmark{1}\IEEEauthorrefmark{3},
Fedelucio Narducci\IEEEauthorrefmark{1},
Azzurra Ragone\IEEEauthorrefmark{2},
Michelantonio Trizio\IEEEauthorrefmark{3},
Claudio Bartolini\IEEEauthorrefmark{3}
}

\IEEEauthorblockA{\IEEEauthorrefmark{1}Polytechnic University of Bari, Bari, Italy\\
Email: \{andrea.lops, fedelucio.narducci\}@poliba.it}

\IEEEauthorblockA{\IEEEauthorrefmark{2}University of Bari, Bari, Italy\\
Email: azzurra.ragone@uniba.it}

\IEEEauthorblockA{\IEEEauthorrefmark{3}Wideverse, Bari, Italy\\
Email: \{andrea.lops, michelantonio.trizio, claudio.bartolini.consultant\}@wideverse.com}
}

\maketitle

\begin{abstract}
Unit testing is an essential but resource-intensive step in software development, ensuring individual code units function correctly. This paper introduces \textsc{AgoneTest}, an automated evaluation framework for Large Language Model-generated (LLM) unit tests in Java. \textsc{AgoneTest} does not aim to propose a novel test generation algorithm; rather, it supports researchers and developers in comparing different LLMs and prompting strategies through a standardized end-to-end evaluation pipeline under realistic conditions.
We introduce the \textsc{Classes2Test} dataset, which maps Java classes under test to their corresponding test classes, and a framework that integrates advanced evaluation metrics, such as mutation score and test smells, for a comprehensive assessment.
Experimental results show that, for the subset of tests that compile, LLM-generated tests can match or exceed human-written tests in terms of coverage and defect detection. Our findings also demonstrate that enhanced prompting strategies contribute to test quality.
\textsc{AgoneTest} clarifies the potential of LLMs in software testing and offers insights for future improvements in model design, prompt engineering, and testing practices.
\end{abstract}


\section{Introduction}

Software testing is a critical step in the software development lifecycle, essential for ensuring code correctness and reliability. Unit testing, in particular, verifies the proper functioning of individual code units.
However, designing and building unit tests is a costly and labor-intensive process that requires significant time and specialized skills \cite{DBLP:conf/issre/DakaF14}. Automating this process is an active area of research and development.

Automated tools for generating unit tests can reduce the workload of test engineers and software developers. These tools typically use static code analysis methods to generate test suites. For example, EvoSuite \cite{fraser2011evosuite}, a popular tool that combines static code analysis with evolutionary search, has demonstrated the ability to achieve adequate coverage.

Large Language Models (LLMs), efficiently exploited in various aspects of software development, could also handle the automatic generation of unit tests. Several empirical studies on LLMs have highlighted their ability to generate tests for simple scenarios, often limited to single methods \cite{guilherme2023initial, schafer2023empirical, siddiq2024using, yuan2023no}. Though directionally useful, these explorations often focus on independent, small-scale test units and rely on manual integration into projects, providing a limited view of LLM performance in real-world software development scenarios \cite{tang2024chatgpt, yuan2023no}. This manual process restricts the number of tests that can be executed and reduces overall efficiency.

To address these gaps, we have developed a framework explicitly focused on the evaluation of unit test suites generated by LLMs. Rather than proposing a novel generation method, our contribution lies in providing an end-to-end pipeline that standardizes how LLM-based test suites can be assessed in realistic software projects.
A simple use case illustrates how \textsc{AgoneTest} can be applied in a real-world scenario. Imagine a developer or a researcher who needs to evaluate which LLM and prompting strategy performs best for generating unit tests. Doing this manually would require repeated project setup, test execution, and metric collection, making the process slow and error-prone. With \textsc{AgoneTest}'s standardized end-to-end pipeline, the developer can automate the workflow and directly compare LLMs under different prompting strategies. The framework produces reliable and reproducible metrics, revealing, for instance, that one model generates more compilable tests while another achieves higher coverage. In this way, \textsc{AgoneTest} turns ad hoc experimentation into systematic benchmarking.
Our approach focuses on class-level test code evaluation, which is closer to real-world practices as it covers method interactions and shared state, reducing code redundancy \cite{DBLP:journals/jss/GranoPNLG19}.

For instance, a simple \texttt{ItemManager} class with two methods: \texttt{addItem()} and \texttt{getItemCount()}, illustrates this point. 
A method-level test for \texttt{getItemCount()} in isolation might only verify its behavior on an empty list, ignoring how the state changes when new items are added. In contrast, a class-level test naturally exercises the interaction between \texttt{addItem()} and \texttt{getItemCount()}, ensuring that the internal state is updated consistently across method calls. 
This example highlights three key benefits of class-level testing:  
\begin{itemize}
    \item \textbf{Reduction of redundancy}: common setup code (e.g., creating and initializing an object) can be reused across multiple methods.  
    \item \textbf{Coverage of complex interactions}: tests verify how methods behave together, capturing issues that method-level tests may miss.
    \item \textbf{Holistic view of class behavior}: class-level tests reveal whether the class as a whole fulfills its intended responsibilities, beyond the correctness of individual methods.  
\end{itemize}

In this work, we introduce \textsc{AgoneTest}, an automated evaluation system for LLM-generated unit tests. \textsc{AgoneTest} integrates project setup, context extraction, execution of generated tests, and quality measurement using standard metrics. Leveraging the \textsc{Methods2Test} dataset \cite{tufano2022methods2test}, we developed a new dataset specifically aimed at comparing human-written tests with those produced by LLMs.

The \textbf{main contributions} of our work are as follows:
\begin{itemize}
    \item \textbf{\textsc{AgoneTest}}\footnote{\url{https://anonymous.4open.science/r/classes2test}\label{note1}}: we designed and developed a closed-loop, highly automated software system supporting the process of assessing LLM-generated unit tests, with automated project setup, prompt integration, and metrics computation.
    \item A unified framework for comprehensive evaluation of a variety of LLMs and relative prompting types and prompt schemata in the task of developing unit tests, and a set of metrics and test smells to assess the quality of the generated test suites;
    \item \textsc{\textbf{Classes2Test}}\footref{note1}: An annotated open source Java project dataset extending \textsc{Methods2Test} \cite{tufano2022methods2test}, which maps classes under test to their related test classes. This extended dataset makes it possible to assess the test performance of an LLM on a more complex scope (the entire class) than the single method.
\end{itemize}

The paper is organized as follows. Section~\ref{sec:background&related} sets the background and highlights differences between our work and related work. 
Section~\ref{sec:overview} gives an overview of \textsc{AgoneTest} and its modules, detailing their functional scope. 
 Then, Section~\ref{sec:in_practise} showcases how \textsc{AgoneTest} is applied in practice through an end-to-end example. Section~\ref{sec:evaluation} presents the Research Questions (RQs) guiding the experiment and the evaluation settings, while Section~\ref{sec:Discussion} answers the research questions and discusses the results. Section~\ref{sec:limitations} discusses the limitations of our approach, and Section~\ref{sec:conclusion} concludes the paper, outlining potential directions for future work.

 \section{Background and Related Work}
\label{sec:background&related}

\subsection{Unit Test Generation}
Unit test generation is the automated process of creating test cases for individual software components, such as functions, methods, or modules. These test cases are used to independently verify the correct functioning of each unit.

Present techniques employ randomness-based \cite{csallner2004jcrasher, pacheco2007feedback}, constraint-based \cite{ma2015grt, sakti2014instance}, or search-based approaches \cite{andrews2011genetic, derakhshanfar2022basic}. The core idea behind these methods is to transform the problem into one that can be solved mathematically. For example, search-based techniques convert testing into an optimization problem, to generate unit test cases \cite{tonella2004evolutionary}. Consequently, the objective of these techniques is to generate all potential solutions and then select those that achieve better code coverage. 

Tools like JUGE have been proposed to evaluate and compare Java unit test generators comprehensively. JUGE provides an infrastructure for benchmarking various Java unit test generation tools by measuring their performance across standard datasets and metrics, enabling a fair and systematic evaluation \cite{devroey2023juge}. While JUGE focuses on benchmarking and comparing classical tools, \textsc{AgoneTest} not only allows unit test generation using multiple tools and techniques, including LLM-based approaches, but also integrates automated assessment into a single, unified framework, making it a practical solution for real-world software development workflows.

EvoSuite \cite{fraser2011evosuite} generates unit tests for Java by applying search-based algorithms that evolve candidate test suites toward coverage objectives such as line and branch coverage. The tool evaluates test fitness iteratively through variation, selection, and optimization, and it automatically produces JUnit test cases along with detailed reports on metrics like code coverage and mutation score. Despite its popularity, EvoSuite has notable limitations. It often produces tests that lack clarity and readability \cite{grano2018empirical}, which hinders their practical usefulness. Moreover, EvoSuite only supports Java 9 or earlier versions and appears to be no longer actively maintained, restricting its applicability to modern Java projects. 
In contrast, \textsc{AgoneTest} overcomes this barrier by supporting all Java LTS versions.

Randoop \cite{pacheco2007feedback} is a feedback-directed random test generator for Java programs. It works by repeatedly selecting sequences of method and constructor invocations, executing them, and using the observed program behavior to guide further generation. Randoop is lightweight and can quickly produce large numbers of test cases that expose common programming errors such as null dereferences, assertion violations, or unhandled exceptions~\cite{DBLP:conf/fsen/PaydarA19}. However, the generated tests often require additional curation: they may include redundant or uninformative assertions, depend on specific runtime states, or fail to integrate smoothly into existing build systems. Unlike \textsc{AgoneTest}, which emphasizes project-level integration and automated assessment, Randoop is typically applied at the level of individual classes with per-class configuration, making large-scale, unattended evaluation less practical.

\subsection{Large Language Models for Test Generation}
Since the emergence of LLMs, they have been used for test suite generation. The first techniques exploiting LLMs were thought of as solutions to neural machine translation problems \cite{nie2023learning, tufano2020unit}. Such approaches work by translating from the primary method to the appropriate test prefix or test assertion while also fine-tuning the LLMs using the test generation dataset. For instance, AthenaTest \cite{tufano2020unit} optimizes BART \cite{chipman2010bart} using a test generation dataset in which the source is the primary method along with its corresponding code context, and the result is the complete test case.

AthenaTest focuses mainly on generating method-level tests by fine-tuning a single model, while \textsc{AgoneTest} shifts the focus to the generation of class-level tests. Our approach makes it possible to use up-to-date LLMs and not constrain prompt design (our prompts can be customized), thereby handling more complex, real-world scenarios.

In light of the rapid evolution of instruction-tuned LLMs, the proliferation of methods for generating tests is on the rise, exploiting guided LLMs through appropriate prompts, as opposed to model fine-tuning \cite{deng2023large, xia2024fuzz4all}. Several proposals for evaluating LLMs in test suite generation have emerged. For example, \textsc{ChatTester} \cite{yuan2023no} proposes a tool for evaluating and improving LLM-generated tests based on ChatGPT.

ChatTester focuses on improving and evaluating tests generated by a specific LLM (ChatGPT), but requires human intervention to evaluate the generated code and does not provide an evaluation of class-level tests on multiple LLMs. \textsc{AgoneTest} provides support instead for a variety of LLMs and evaluates each LLM's performance on a wide range of real-life Java projects. \textsc{TestPilot} \cite{schafer2023empirical} is also focused on generating and improving tests using LLMs on JavaScript code. Although TestPilot performs an automated evaluation, it lacks wider applicability to projects other than the 25 repositories it considers in the work provided as reference here. \textsc{AgoneTest} offers far broader applicability by using a dataset of 9,410 GitHub repositories, and automatically integrating test libraries into them. \textsc{Cedar} \cite{nashid2023retrieval} instead proposes a prompt construction strategy based on \textit{few-shot} learning\cite{brown2020language} and the Codex model\footnote{\href{https://openai.com/index/openai-codex/}{https://openai.com/index/openai-codex/}} to generate tests. 

Cedar uses a specific prompt construction strategy but does not incorporate a structured mechanism to evaluate multiple LLMs and prompt types in a unified framework. \textsc{AgoneTest} provides this by allowing the integration and evaluation of various prompt engineering techniques and LLMs, offering a more holistic approach to test generation. Guilherme and Vincenzi \cite{guilherme2023initial} use \texttt{gpt-3.5-turbo} in analyzing the impact of variation in model hyperparameters. 

The study by Guilherme and Vincenzi presents an initial assessment but lacks automation in evaluating comprehensive test quality metrics like mutation score and test smells. \textsc{AgoneTest} goes a step further by automating these evaluations, integrating advanced metrics to provide a deeper analysis of the generated tests. Siddiq et al. \cite{siddiq2024using} offer a new proposal for evaluating tests generated using common datasets and experimenting with the use of new metrics \cite{palomba2016diffusion}. 

Although Siddiq et al. use Test Correctness (but not mutation score) on top of all the metrics that \textsc{AgoneTest} uses, their approach does not fully automate the test generation-execution-evaluation loop or focus on class-level tests. \textsc{AgoneTest} fills this gap by providing end-to-end automation and focusing on generating and evaluating complex, class-level test suites.

A number of other LLM-based test generation tools have been proposed, but they suffer from limitations that make them unsuitable for systematic comparisons with \textsc{AgoneTest}. 
For instance, \textsc{A3Test}~\cite{DBLP:journals/infsof/AlagarsamyTA24}, \textsc{ChatUniTest}~\cite{chen2024chatunitest}, \textsc{TestSpark}~\cite{DBLP:conf/icse/SapozhnikovOPKD24}, and \textsc{ProjectTest}~\cite{DBLP:journals/corr/abs-2502-06556} proved unusable in our experiments due to critical issues such as non-functional setup scripts, token limit errors even on small inputs, inability to generate a single compilable test without extensive manual intervention, and disregard for project-specific dependencies. Similarly, our analysis of \textsc{TestBench}~\cite{DBLP:journals/corr/abs-2409-17561} revealed that its provided codebase was not functional out-of-the-box, producing immediate compilation errors and lacking the necessary requirements for systematic evaluation. 

Other tools differ in goals or scope: for example, \textsc{TestGen-LLM}~\cite{DBLP:conf/sigsoft/AlshahwanCFGHHM24} requires manual validation for each test, thus breaking the automated loop that is central to \textsc{AgoneTest}. \textsc{MuTAP}~\cite{DBLP:journals/infsof/DakhelNMKD24} uses mutation score to improve prompt engineering, whereas \textsc{AgoneTest} employs mutation score as a final quality metric for assessing the robustness of the entire generated test suite. Finally, tools such as \textsc{Qodo Cover}\footnote{\url{https://github.com/qodo-ai/qodo-cover}} and \textsc{UTBot}\footnote{https://github.com/UnitTestBot} focus primarily on the generation process and have their target LLMs hard-coded, preventing comparative studies across multiple models and prompting strategies.

\subsection{Limits of Current Approaches in Applying LLMs to Unit Test Generation}\label{subsec:limits_current}
While promising, current approaches in applying LLMs to unit test generation exhibit several limitations:

\paragraph{Limited Scope} Current methods for assessing how useful LLMs are in test code generation are mostly limited to the generation of code segments, rather than whole modules or components (e.g., whole classes in Java). Consequently, until very recently, the research community lacked mature and widely adopted datasets for evaluating class-level test generation.
Some early benchmarks have now started to appear, including \textsc{TestGenEval}~\cite{DBLP:conf/iclr/JainSR25}, \textsc{TestBench}, and \textsc{ProjectTest}. These works confirm the importance of class- and project-level testing, but they differ from our contribution in scope and maturity: \textsc{TestGenEval} focuses on single classes and limited metrics, \textsc{TestBench} provides a prototype implementation with limited usability, and \textsc{ProjectTest} explores project-level mappings but lacks full automation.

Other studies often provide only punctual and anecdotal evaluations of the generated results, evaluating LLMs in the task of generating tests only at the method level or in contexts limited to sections of code \cite{tufano2020unit, yuan2023no, schafer2023empirical}.

\paragraph{Lack of Automation} While some prior works (e.g., \textsc{TestGenEval}, \textsc{TestBench}, \textsc{ProjectTest}) have started exploring aspects of automated evaluation, none provide a fully automated end-to-end framework that integrates project setup, test generation, execution, and metric computation in a reproducible pipeline. \textsc{AgoneTest} fills this gap by offering an extensible evaluation framework rather than a one-off benchmark.

\paragraph{Subjective Choice of Prompts} In most cases, the choice of prompts to get LLMs to generate testing code remains subjective. There is no thorough evaluation of alternate prompting techniques compared to those initially proposed, leaving room for further exploration and optimization in prompt engineering. \cite{siddiq2024using, nashid2023retrieval, xia2024fuzz4all}.

\begin{table*}[t]
\caption{Comparison between \textsc{AgoneTest} and recent class-/project-level test–generation benchmarks/frameworks.}
\label{tab:rw_compare}
\centering
\resizebox{\textwidth}{!}{
\begin{tabular}{l|l|l|l|c|c|l|c|c}
\thead{Work} & \thead{Language(s)} & \thead{Input scope} & \thead{Primary aim} & \thead{End-to-end\\automation\\(setup$\rightarrow$metrics)} & \thead{LLM/\;prompt\\configurable} & \thead{Reported metrics} & \thead{Build/\;deps\\integration} & \thead{Post-gen\\repair/\;compile\\boost}\\
\hline
\textsc{AgoneTest} (this work) & Java & \makecell[l]{Class-level within\\full project} & \makecell[l]{Evaluation framework\\(multi-LLM,\;multi-prompt)} & \textcolor{ForestGreen}{Yes} & \textcolor{ForestGreen}{Yes} & \makecell[l]{Line,\;Branch,\;Method Coverage\\Mutation score,\\18 Test smells} & \textcolor{ForestGreen}{Yes} (Maven/Gradle) & \makecell[l]{\textcolor{ForestGreen}{Yes} (enhanced\\prompting for\\imports/paths)}\\
\hline
\textsc{TestGenEval}~\cite{DBLP:conf/iclr/JainSR25} & \makecell[l]{Primarily\\Python} & \makecell[l]{Single class\\(not full project)} & \makecell[l]{Benchmark of\\generation quality} & \textcolor{BrickRed}{No} & \textcolor{BrickRed}{No} & \makecell[l]{Line Coverage,\\Mutation score} & \textcolor{BrickRed}{N/A} (no project build) & \textcolor{BrickRed}{No}\\
\hline
\textsc{TestBench}~\cite{DBLP:journals/corr/abs-2409-17561} & Java & \makecell[l]{Class-level\\(repo subset)} & \makecell[l]{Benchmark/analysis\\of LLM tests} & \textcolor{BrickRed}{No} & \textcolor{BrickRed}{No} & \makecell[l]{Line,\;Branch,\;Method Coverage,\;\\Mutation score} & \textcolor{BrickRed}{No} & \textcolor{BrickRed}{No}\\
\hline
\textsc{ProjectTest}~\cite{DBLP:journals/corr/abs-2502-06556} & Java & \makecell[l]{Project-level\\mapping focus} & \makecell[l]{Project-level\\benchmarking} & \textcolor{BrickRed}{No} & \textcolor{BrickRed}{No} & \makecell[l]{Line,\;Branch,\;Method Coverage,\;\\Mutation Score} & \textcolor{BrickRed}{No} & \textcolor{BrickRed}{No}\\
\end{tabular}
}
\end{table*}

Table \ref{tab:rw_compare} summarizes the characteristics of these works and compares them with ours.

\paragraph{Contribution of \textsc{AgoneTest}} 
In contrast to these limitations, \textsc{AgoneTest} provides three distinctive features that set it apart from prior work. 
First, it offers full automation, covering the entire loop from repository setup to prompt instantiation, test generation, execution, and quality assessment, without requiring manual steps. 
Second, it explicitly targets class-level testing, which more closely reflects real-world development practices by capturing interactions between methods and shared state within classes, going beyond the common focus on isolated methods. 
Finally, \textsc{AgoneTest} is designed as an extensible framework, allowing researchers and practitioners to integrate new LLMs, prompt strategies, and datasets with minimal effort, thus serving as a reusable infrastructure rather than a one-off experiment.

\section{Overview of \textsc{AgoneTest}}
\label{sec:overview}
The term \textit{agone}, originating from ancient Greece and Rome, signified a contest wherein philosophers debated their ideas, with the audience determining the victor. We adopt the term \textit{agone} metaphorically to represent the competitive evaluation of LLMs and their respective prompting strategies within an arena aimed at generating optimal unit test suites. \textsc{AgoneTest} determines the optimal strategies based on standard test quality metrics, see Sec.~\ref{subsection:test-evaluation}.

\textsc{AgoneTest} is designed to provide software testers with a system for generating and assessing unit tests. This assessment focuses on key metrics such as code coverage, defect detection rate, and the presence of known test smells, thereby offering a comprehensive assessment of test suite quality.

\textsc{AgoneTest} operates on the principle that the evaluation of LLMs in the task of generating high-quality unit tests can be performed through the collaboration of test engineers and data scientists (or prompt engineers). However, in practice, a single experienced test engineer familiar with generative AI can perform both roles, allowing the focus to be only on defining new prompt types and the comparison of LLMs. This is the persona that we evoke when we refer to the \textsc{AgoneTest} user (alternatively, ``the test engineer”) in the remainder of this paper. 

\begin{figure}[ht]
    \centering
    \includegraphics[width=0.5\textwidth]{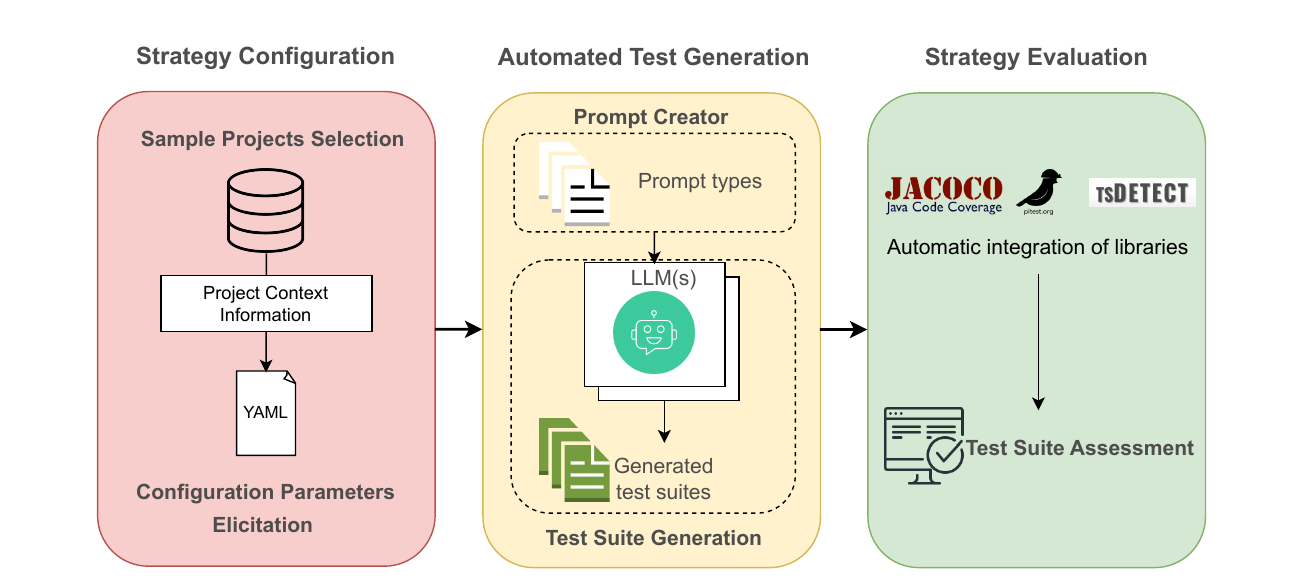}
    \caption{Overview of \textsc{AgoneTest} framework}
    \label{fig:overview}
\end{figure}

\Cref{fig:overview} provides a high-level diagram of the architecture of \textsc{AgoneTest}, showing the operating modules that streamline the test generation and evaluation process.
The framework can be described as follows:

\subsubsection{Strategy Configuration}
\paragraph{Sample Projects Selection} As an initial configuration step, the user chooses which repositories to generate test suites for. This initial phase leverages a comprehensive dataset of open-source Java repositories, which we contribute to the community.
\paragraph{Configuration Parameters Elicitation} In this phase, configuration parameters are elicited from the selected repositories (e.g., the project Java version, the used testing framework, etc.) and processed to create prompt templates that will be handed over to the LLMs.
\subsubsection{Automated Test Generation}
\paragraph{Prompt Creator} During this phase, the prompt templates built in the previous phase are fully instantiated and then used to generate unit test suites in the next step.
\paragraph{Test Suite Generation} Here, \textsc{AgoneTest} orchestrates the interaction with the user's selected LLMs, feeding them the instantiated prompts to produce the unit test code. Each LLM generates test classes, which are then automatically integrated into the project structure.

\subsubsection{Strategy Evaluation}
\paragraph{Test Suite Assessment} This phase assesses the quality of the test suites by computing various metrics and identifying (if any) test smells.
This assessment enables a detailed analysis of the effectiveness and quality of the automated test generation strategies.\\
In the following, we describe each phase of the process in detail.

\subsection{Sample Projects Selection: \textsc{Classes2Test} dataset}
We note that the \textsc{AgoneTest} tool can import and process any Java industrial project. 
For the purpose of evaluating our system, we built a comprehensive and annotated dataset of open-source Java repositories from GitHub, which we leverage in this phase. Unlike popular datasets in the literature, our dataset enables the generation and validation of unit tests at the Java class level, rather than at the individual method level. 
To collate and annotate our dataset, we built on \textsc{Methods2Test} \cite{tufano2022methods2test}. 
The \textsc{Methods2Test} dataset was originally built from an initial pool of 91,385 GitHub repositories. To ensure realism and maintainability, only projects satisfying strict criteria were retained, yielding approximately 10\% of the initial corpus (9,410 repositories). The filtering procedure excluded repositories that:
\begin{itemize}
    \item ``had not been updated in the past five years,"
    \item ``were forks or duplicates,"
    \item ``failed to compile successfully after dependency resolution,"
    \item ``or did not contain a sufficient number of paired method–test mappings."
\end{itemize}

This process ensured that the resulting dataset captures actively maintained, compilable, and representative real-world projects.
For \textsc{Classes2Test}, we extracted information not only on classes and their mapped test classes under test but also on test frameworks and Java versions used across projects: JUnit~4 (55\%), JUnit~5 (41\%), and others (4\%, including TestNG). Regarding Java versions, the majority of repositories target Java 11 (42\%), followed by Java 17 (25\%), newer LTS versions such as Java 21 (18\%), and Java 8 (14\%). 
We chose \textsc{Methods2Test} as the starting point, as it contains not only the test methods to be tested, but also the corresponding test methods written and validated by humans. Human-written tests are a valid evaluation benchmark to evaluate the effectiveness of different LLMs in building a test suite.

To build the \textsc{Classes2Test} dataset, we extracted all references to the open-source repositories present in \textsc{Methods2Test} in order to map the Java classes, referred to as \textit{classes under test}, to their corresponding test classes.

Here is the process we followed to create \textsc{Classes2Test}:
\begin{enumerate}
    \item Extract the repository reference, GitHub URL, and selected branch;
    \item Select the classes considered in \textsc{Methods2Test};
    \item Clone the repository and save the commit hash;
    \item Map and save the classes under test along with their respective test classes.
\end{enumerate}

To map classes under test to their corresponding test classes in \textsc{Classes2Test}, we adopted a conservative two-step procedure. First, candidate test classes are identified by common Java naming conventions (e.g., \texttt{MyClassTest}, \texttt{TestMyClass}) within the \texttt{src/test/java} folder, also considering mirrored package structures. Second, these candidates are statically validated through AST analysis to confirm that they actually exercise the class under test, checking for imports, constructor calls, method invocations, or mocking references. Only pairs with sufficient structural evidence are retained. In cases where a test class references multiple classes under test, we compute an evidence ratio and keep the mapping only if one class under test dominates ($\geq$60\%of references); otherwise, the test class is discarded. Ambiguous or conflicting mappings (e.g., two classes under test pointing to the same test class with comparable evidence) are excluded to ensure precision. In the rare cases where a candidate test class exhibited comparable structural evidence for multiple classes under test (i.e., no dominant mapping could be established), we conservatively excluded the mapping altogether to avoid spurious links. This situation occurred only in 3 instances out of the entire corpus ($\approx$0.002\% of all analyzed classes), and the affected pairs were removed from \textsc{Classes2Test}. This process was inspired by the methodology used in \textsc{Methods2Test}.

\begin{table}
\centering
\caption{\textsc{Classes2Test} Dataset Characteristics}
\resizebox{1\linewidth}{!}{
\begin{tabular}{l|r}
Characteristic & Value\Tstrut\Bstrut\\
\hline
\# Test Classes & 147,473\Tstrut\\
\# Unique Repositories & 9,410 \\
Average Lines of Code per Class & 1,178 (IQR: 420–1,960) \\
Average Cyclomatic Complexity per Class & 55.3 (IQR: 18–92) \\
Test Framework Distribution & JUnit~4 (55\%), JUnit~5 (41\%), Other (4\%) \\
Java Version Distribution & 8 (14\%), 11 (42\%), 17 (25\%), 21+ (18\%) \\
\end{tabular}
}
\label{tab:dataset_characteristics}
\end{table}

The resulting dataset contains 147,473 test classes extracted from 9,410 unique repositories. A summary of the dataset's characteristics is shown in Table \ref{tab:dataset_characteristics}.

\subsection{Configuration Parameters Elicitation}
\label{sub:user_input}
Before unit test generation can begin, the system extracts some parameters from the projects selected in the previous step. These parameters are then fed into the module that instantiates prompts and selects LLMs. 
To query the model under examination, various prompting techniques are available and can be chosen by the test engineer \cite{sahoo2024systematic}.

The configuration parameters include: 
\begin{itemize}
    \item \texttt{class\_under\_test}: This variable contains the Java class for which the test suite must be generated;
    \item \texttt{testing\_framework}: This variable provides the name and version of the project's testing framework (e.g., JUnit 4), directly extracted from the project during execution;
    \item \texttt{java\_version}: This variable allows you to retrieve the version of Java that the project uses.
    \item \texttt{example\_class\_under\_test \& example\_test\_class}: These variables contain an example class under test and the corresponding test class extracted from a reference repository, useful to provide an example to the LLM if one wants to use the few-shot prompting technique;
    \item \texttt{example\_testing\_framework \& example\_java\_version}: These variables provide the information about the example repo.
    The example consists of a class under test and a test class extracted from an open sample repository\footnote{\href{https://github.com/junit-team/junit5-samples/blob/main/junit5-jupiter-starter-maven/src/test/java/com/example/project/CalculatorTests.java}{https://github.com/junit-team/junit5-samples}}. Future iterations of \textsc{AgoneTest} will automate exemplar selection to reduce bias and better reflect real-world scenarios.
\end{itemize}
See Sec.~\ref{subsec:LLMs-selection} for an example of a real implementation. 

\subsection{Prompt Creation}
\label{sub:prompt_creation}

In this phase, the prompt templates described in the previous phases are fully instantiated to create viable prompts to guide the LLM in generating unit tests.
We populate the user-supplied prompt types by replacing the variables outlined in Sec.~\ref{sub:user_input}.

It has to be noted that, in order to make sure our experiments and findings are reproducible, we prepared \textsc{Classes2Test} by saving the commit hashes of the repositories used as sources. This allows \textsc{AgoneTest} to consistently extract information such as the Java version used, the type of test framework (e.g., JUnit), and its version.

Unlike previous approaches to creating unit testing with LLMs that require human intervention to input code information \cite{yuan2023no, guilherme2023initial}, \textsc{AgoneTest} automates the process to a far greater degree. \textsc{AgoneTest} employs ElementTree \cite{garabik2005processing} and a parser to read and modify the Maven and Gradle build (see Sec.~\ref{subsection:test-evaluation}). It analyzes the libraries present and the Java version used in each build system.
The prompting engine of \textsc{AgoneTest} is designed to be extensible and flexible. Beyond the zero-shot and few-shot strategies illustrated in our examples (see Sec.~\ref{subsec:LLMs-selection}), the YAML-based configuration supports custom variables that users can freely define to adapt prompts to new scenarios. This design makes it straightforward to incorporate advanced prompting techniques such as multi-shot prompting (with multiple test pair examples) or retrieval-augmented generation~\cite{DBLP:conf/nips/LewisPPPKGKLYR020} (injecting external context). The engine does not hard-code prompt types, but instead provides a generic schema that researchers and practitioners can extend for novel experimental setups.

\subsection{Test Suite Generation}
At this point in the process, we have everything we need for the selected LLMs to generate test suites for each class under test of the project. To ensure each model has an appropriate number of tokens, we use tiktoken\footnote{\href{https://github.com/openai/tiktoken}{https://github.com/openai/tiktoken}}, a BPE tokenizer \cite{berglund2023formalizing}, to evaluate the token count in the prompt. When the token limit of a target model is exceeded, \textsc{AgoneTest} provides configurable fallback strategies rather than failing silently. By default, the system notifies the user of the excess and the number of tokens required. The user can then:
(i) truncate the input, for instance by shortening comments or omitting less relevant methods;
(ii) manually adjust the prompt template to reduce verbosity or context.

We remark that \textsc{AgoneTest} allows users to automatically choose and evaluate a wide range of LLMs. This capability is provided by the open-source LiteLLM library\footnote{\href{https://github.com/BerriAI/litellm}{https://github.com/BerriAI/litellm}}, which facilitates communication with more than 100 models\footnote{\href{https://docs.litellm.ai/docs/providers}{https://docs.litellm.ai/docs/providers}} using a standard interaction based on the OpenAI API format\footnote{\href{https://platform.openai.com/docs/guides/text-generation/chat-completions-api}{https://platform.openai.com/docs/guides/text-generation/chat-completions-api}}. Integration is made easier by LiteLLM, which translates inputs to satisfy the unique endpoint needs of each provider. This is crucial in today's environment, where the absence of standard API specifications for LLM providers makes it challenging to incorporate several LLMs into projects.

After invoking the LLM, \textsc{AgoneTest} selects relevant information from the LLM's answer (i.e., the generated test class). This step is crucial for automating the entire process, since LLMs can provide detailed descriptions or explain how the code should be structured without actually generating it \cite{chen2024chatunitest}. In this component, \textsc{AgoneTest} removes unnecessary parts (like outline descriptions) and creates a new file to integrate the test class into the project.

\subsection{Test Suite Assessment}
\label{subsection:test-evaluation}
Here, we evaluate the quality of the test suite according to the quality metrics described in the following. Unlike previous methods \cite{siddiq2024using, yuan2023no} that require manual or partial automation, our framework provides a fully automated evaluation of the generated tests through systematic integration of evaluation metrics, representing a distinct step forward in automated evaluation.
It is important to note that this component is separate from the experimental evaluation discussed later. Instead, it serves as an additional tool provided by \textsc{AgoneTest} to assist engineers in assessing the quality of the generated tests.

\subsubsection{Code Coverage} we calculate several coverage metrics, specifically: Line coverage, Method coverage, and Branch coverage \cite{aniche2022effective}. To measure code coverage in test suite execution, we integrated in \textsc{AgoneTest} the JaCoCo\footnote{\href{https://www.jacoco.org/jacoco/index.html}{https://www.jacoco.org/jacoco/index.html}} library.

\subsubsection{Defect detection rate} to measure the robustness of the test suite, we use mutation score. The mutation score evaluates the capability of tests to detect syntactic changes (mutants) artificially introduced in the code. This provides a widely used and reliable proxy for test suite robustness, as shown in prior literature \cite{DBLP:journals/ac/PapadakisK00TH19}.

\subsubsection{Test Smells\texorpdfstring{\cite{palomba2016diffusion}}{}}
we decided to identify test smells in the generated test suite as proxy indicators of potential quality and maintainability issues in the test code. Although test smells do not directly measure functional correctness, their presence often correlates with problematic patterns that could negatively affect the effectiveness and readability of the generated tests. To identify these smells in the generated test suite, we integrate the library \textsc{tsDetect} \cite{peruma2020tsdetect}.
\textsc{AgoneTest} determines whether the following test smells are present in the code: Assertion Roulette (AR) \cite{van2001refactoring}; Conditional Test Logic (CTL) \cite{meszaros2003test}; Constructor Initialization (CI) \cite{peruma2019distribution}; Default Test (DT); Duplicate Assert (DA) \cite{peruma2019distribution}; Eager Test (EA) \cite{van2001refactoring}; Empty Test (EM) \cite{peruma2019distribution}; Exception Handling (EH) \cite{peruma2019distribution}; General Fixture (GF); Ignored Test (IT) \cite{peruma2019distribution}; Lazy Test (LT) \cite{van2001refactoring}; Magic Number Test (MNT) \cite{meszaros2003test}; Mystery Guest (MG); Redundant Print (RP) \cite{peruma2019distribution}; Redundant Assertion (RA) \cite{peruma2019distribution}; Resource Optimism (RO) \cite{peruma2019distribution}; Sensitive Equality (SE) \cite{van2001refactoring}; Sleepy Test; Unknown Test (UT) \cite{peruma2019distribution}. 

After adding the necessary libraries, \textsc{AgoneTest} runs a build and test to ensure that there are no compilation errors. 
The assessment phase of the test suite of our process presents a high degree of automation, as we describe below.

In this phase, \textsc{AgoneTest} automatically includes these libraries into the project. For each run, \textsc{AgoneTest} checks the configuration files of the supported build systems (Maven and Gradle, Sec.~\ref{sub:prompt_creation}) to determine if the necessary libraries are already present. If they are not, it modifies the configuration to add the required dependencies.

\textsc{AgoneTest} provides a high degree of automation, for example, in its handling of the PiTest library. Specifically, if the repo uses the JUnit 5 test framework, an additional library, ``pitest-junit5-plugin”, is required. Utilizing information extracted from the repo in the Prompt Creation module (Sec.~\ref{sub:prompt_creation}), \textsc{AgoneTest} automatically identifies the test framework in use and adds this dependency without any human intervention.

\textsc{AgoneTest} generates a report with the results of the quality metrics computed for the LLM-generated test suite. 
To achieve this, the tool automatically retrieves detailed information from the reports produced by the libraries, compiling this data for each class within each project.

\paragraph*{Failure handling.}
All coverage and mutation metrics are computed on the full evaluation set by assigning a value of $0$ to non-compiling generations (Sec.~\ref{sec:eval-protocol}). Test smell counts are build-penalized in the same way (non-compiling $\rightarrow 0$ contribution). We also report the compilation rate to contextualize the aggregated scores.




It is important to note that \textsc{AgoneTest} is not limited to the \textsc{Classes2Test} dataset. 
Thanks to its modular architecture, the framework can seamlessly integrate with alternative benchmarks such as \textsc{Defects4J}~\cite{DBLP:conf/issta/JustJE14} or any other curated corpus of Java projects. 
Researchers need only adjust the configuration files to extend the evaluation to new projects.

\section{\textsc{AgoneTest} in practice}
\label{sec:in_practise}
In this section, we will demonstrate how \textsc{AgoneTest} operates in practice by describing an end-to-end run of a practical example.

We will skip the repository selection phase in our account and move straight to the configuration phase, which concerns LLM selection and prompt specification. Then, we will exemplify how the results are presented back to the user for further analysis.

\subsection{Configuration}\label{subsec:LLMs-selection}
As described in Sec.~\ref{sub:user_input}, \textsc{AgoneTest} utilizes a YAML\footnote{\url{https://yaml.org/}} file as input, where it is possible to specify information related to two elements: \texttt{llms} and \texttt{prompts}. 
The YAML file represented in the Listing \ref{listing:1} is an example configuration for \textsc{AgoneTest} that will be used in Sec.~\ref{sec:in_practise}.

\begin{lstlisting}[
  caption=Setup of the YAML configuration file: setting of variables for two different LLMs and two different prompts.,
  label=listing:1,
  breaklines=true,
  language=yaml,
  basicstyle=\scriptsize,
]
llms:
- model: gpt-4o-mini
  temperature: 0
- model: gemini-1.5-pro
  temperature: 0
- model: llama3.1:70b
  temperature: 0
prompts:
- name: zero-shot
  value:
  - role: system
    content: You are provided with Java class. Create a test class that fully tests the proposed Java class using the project information for imports. Reply with code only, do not add other text that is not code.
  - role: user
    content: "The project uses {{testing_framework}} and Java {{java_version}} and Java class is:  \n<code>\n {{class_under_test}}\n</code>"
- name: few-shot
  value:
  - role: system
    content: You are provided with an example with a Java class and its test class. You are then provided with a new Java class. Take a cue from the example and create a test class that fully tests the new proposed Java class. Reply with code only, do not add other text that is not code.
  - role: user
    content: "#Example:\nThe example Java class is:\n<code>\n {{example_java_class}} \n</code>\nThe example test class is: \n<code>\n {{example_test_class}} \n</code>.\nThe Java class you must create the test for is: \n<code>\n{{class_under_test}}\n</code>"
\end{lstlisting}
The LLM is specified by setting the model name in the \texttt{model} field, selecting from the available models supported by LiteLLM\footnote{\href{https://docs.litellm.ai/docs/providers}{https://docs.litellm.ai/docs/providers}}. 
In the \texttt{temperature} field, you can set the temperature at which the model will operate.
 The temperature can have a value between 0 and 2. Higher values (e.g., 2) produce more random outputs, while lower values (e.g., 0) make the outputs more targeted and deterministic \cite{guilherme2023initial}.

The \texttt{prompts} field consists of two sections: \texttt{name} and \texttt{value}. The former is an identifier for labeling the type of prompt (\textit{zero-shot}, \textit{few-shot}, etc.); the latter, \texttt{value}, is an array of message elements of type OpenAI\footnote{\href{https://platform.openai.com/docs/api-reference/chat/create\#chat-create-messages}{https://platform.openai.com/docs/api-reference/chat/create\#chat-create-messages}}. Each individual message includes a \texttt{role} and a \texttt{content}. 

For the \texttt{role} field, there are two possible values:
\begin{itemize}
    \item \texttt{system}: used to instruct the model on the behavior it should adopt.
    \item \texttt{user}:  used to indicate the request for the generation of the test class.
\end{itemize}

Two types of prompts can be specified: 
\begin{itemize}
\item \texttt{zero-shot}: refers to presenting the model with a single instance of a request or task without any previous examples for the model to draw upon \cite{radford2019language}. This method emphasizes the model's ability to comprehend and accurately execute the given task. 
\item \texttt{few-shot}: unlike zero-shot prompting, few-shot prompting involves providing the model with examples demonstrating the expected inputs and outputs \cite{brown2020language}. This technique aids in contextual learning by including examples in the prompt, thereby guiding the model towards improved performance. These examples serve as a conditioning for the actual request, helping the model generate more accurate and relevant responses.
\end{itemize}

This configuration file will instruct \textsc{AgoneTest} to perform the steps described in Sec.~\ref{sub:prompt_creation}: it will fully instantiate template variables, including class under test, test frameworks used by the repos (and versions thereof), and version of the JDK used in the repos.

\subsection{Results presentation}
After performing the generation step, \textsc{AgoneTest} produces a report containing, for each selected LLM and each prompting mode, the quality metrics for the classes under test and the entire project.
Table \ref{tab:cvs} shows an excerpt of the report that also contains human-written test results as they were present in the \textsc{Class2Test} dataset. 
The report provides valuable information on the strengths and weaknesses of each combination of LLM and prompt, showing the average value for each metric. 

In this way, software testers can accurately assess the effectiveness of the LLM in creating usable and effective class-level tests.

\begin{table*}
\caption{Excerpt of the report produced by \textsc{AgoneTest}, showing results for classes under test and selected combinations of LLMs and prompt types. 
The full report is available in the online appendix. Here we display only a representative sample for illustration. 
Test smells are reported from column 10 onwards (see Sec.~\ref{subsection:test-evaluation} for acronyms).}
\resizebox{1\textwidth}{!}{
\begin{tabular}{llrlrrrrrrrrrrrrrrrrrrrr}
\thead{model} & \thead{prompt\\name} & \thead{Project} & \thead{Class under test} & \thead{branch\\coverage} & \thead{line\\coverage}   & \thead{method\\coverage} & \thead{mutation\\score} & \thead{AR} & \thead{CTL} & \thead{CI} & \thead{DA} & \thead{EA} & \thead{EM} & \thead{EH} & \thead{IT} & \thead{LT} & \thead{MNT} & \thead{RP} & \thead{RA} & \thead{RO} & \thead{SE} & \thead{UT} \Tstrut\Bstrut\\\hline
gpt-4o-mini & few-shot & 145256500 & Key & 35.71 & 64.86 & 84.62 & 33.33 & 10 & 0 & 0 &0&10&0&0&0& 28 &10&0&0&0&1&0\Tstrut\\ 
gemini-1.5-pro & few-shot & 145256500 & Key & 57.14 & 81.08 & 85.62 & 33.33 & 10 & 0 & 0 &0&0&0&0&0& 9 &9&5 &0 &0&0 &0\Tstrut\\ 
human & - & 145256500 & Key & 28.57 & 37.84 & 30.77 & 25.0 & 0& 0 & 0 &0&1&0&0&0&0&1&0&0&0&0&0\Tstrut\\ 
llama3.1:70b & zero-shot & 43246449 & FileHandler & - & 100 & 100 & 0 & 1 & 0 & 0 &0&0&0&0&0& 4 &4& 0&0 &1&0 &0\Tstrut\\ 
human & - & 43246449 & FileHandler & - & 100 & 100 & 100 & 1 & 0 & 0 &0&0&0&0&0& 4 &4& 0&0 &1&0 &0\Tstrut\\ 
gpt-4o-mini & zero-shot & 1341207 & HexString & 50 & 92.11 & 100 & 85.71 & 7 & 0 & 0 &0&18&0&0&0& 45 &18&0&0&0&1&0\Tstrut\\ 
gpt-4o-mini & few-shot & 1341207 & HexString & 48.39 & 89.47 & 96.0 & 85.71 & 7 & 0 & 0 &0&17&0&0&0& 38 &17&0&0&0&1&0\Tstrut\\ 
human & - & 1341207 & HexString & 25.81 & 44.74 & 44.0 & 51.43 & 1 & 0 & 0 &0&1&0&1&0&0&1&0&0&0&0&0\Tstrut\\ 
... & ... & ... & ... & ... & ... & ... & ... & ... & ... & ... & ... & ... & ... & ... & ... & ... & ... & ... & ... & ... & ... & ...\\ 
\end{tabular}
}
\label{tab:cvs}
\end{table*}

\section{\textsc{Experiment Setup}}
\label{sec:evaluation}
In this experimental evaluation, our aim is to address the following research questions:
\begin{itemize}
\item \textbf{RQ1: How do the chosen LLMs and prompt types perform in the test case generation task?} We assess this by computing the quality metrics defined in Sec.~\ref{subsection:test-evaluation}.
\item \textbf{RQ2: How frequently do compilation errors occur, and how do they impact the overall compilation success rate?} We study and classify the most common errors that impact the unit test generation process. 
\item \textbf{RQ3: Is there a strategy to increase the compilation success rate?} We study how to increase the compilation success rate. In particular, we define an enhanced strategy to improve the compilation success rate.
\end{itemize}

\subsection{Evaluation protocol: compiled-only averaging (build failures excluded)}
\label{sec:eval-protocol}
Let $N$ be the number of classes under test (since we performed the experiments on the entire \textsc{Classes2Test} dataset, the number of classes will be 147,473). For each class $i$, let $\mathrm{build}_i \in \{0,1\}$ indicate whether the generated tests compile, and let $m_i \in [0,100]$ denote a metric value (e.g., line, branch, method coverage, or mutation score) computed only if $\mathrm{build}_i=1$. Define $N_{\text{comp}}=\sum_{i=1}^{N}\mathrm{build}_i$. We report compiled-only averages (no penalty for non-compiling generations):
\[
\scalebox{1}{ 
  $\displaystyle \overline{m} \;=\; \frac{1}{N_{\text{comp}}} \sum_{i=1}^{N} \mathrm{build}_i \, m_i.$
}
\]

For test smells, let $s_{k,i}\ge 0$ be the count for smell $k$ on class $i$ (defined only if $\mathrm{build}_i=1$). We report compiled-only averages:
\[
\scalebox{1}{ 
  $\displaystyle \overline{s_k} \;=\; \frac{1}{N_{\text{comp}}} \sum_{i=1}^{N} \mathrm{build}_i \, s_{k,i}.$
}
\]
We additionally report the compilation rate $R_{\text{build}}=\frac{1}{N}\sum_{i=1}^{N} \mathrm{build}_i$ to make the influence of failures explicit.

\subsection{Model and Prompt Selection}

We focus on evaluating the performance of LLMs in generating test cases on different types of models. For this purpose, we select \texttt{gpt-4o-mini}, \texttt{gemini-1.5-pro}, and \texttt{llama3.1:70b} for our experimentation. \texttt{gpt-4o-mini} is selected because of its excellent performance in code generation tasks \cite{cui2024webapp1k} and has a number of parameters comparable with the other selected models\footnote{\href{https://openai.com/index/gpt-4o-mini-advancing-cost-efficient-intelligence/}{https://openai.com/index/gpt-4o-mini-advancing-cost-efficient-intelligence}}; \texttt{gemini-1.5-pro} because it is a model that can handle a context of 2 million tokens\footnote{\href{https://developers.googleblog.com/en/new-features-for-the-gemini-api-and-google-ai-studio/}{https://developers.googleblog.com/en/new-features-for-the-gemini-api-and-google-ai-studio}} and allows us to understand whether there are differences in performance based on context size; \texttt{llama3.1:70b} is selected because it is the open-parameter model that has comparable performance with others in the code generation task\footnote{\href{https://ai.meta.com/blog/meta-llama-3-1/}{https://ai.meta.com/blog/meta-llama-3-1/}}. \texttt{llama3.1:70b} is hosted through Ollama\footnote{\href{https://ollama.com/}{https://ollama.com}}and queried locally, while \texttt{gpt-4o-mini} and \texttt{gemini-1.5-pro} models are queried via API.

In addition to choosing LLM, it is crucial to set its temperature parameter. By carefully setting the temperature parameter, users can balance innovation and coherence in text generation, ensuring that the output aligns with their specific task or application requirements.
As shown in Listing \ref{listing:1}, we set the temperature to 0 in our experiment to increase the level of coherence in text generation (and to decrease the level of randomness) and make the different test suites generated reproducible; results are single-run per model/prompt, and \textsc{AgoneTest} exposes temperature for repeated-run studies. Regarding prompt types, we experimented with two of the most popular techniques: zero-shot and few-shot, introduced in Sec.~\ref{subsec:LLMs-selection}.
\begin{table*}
\caption[Average of metrics computed for each model and prompt types used and for human-written tests. In bold, the best results for each metric. Only the test smells with non-zero values are shown in the table.]{Average of metrics computed for each model and prompt types used and for human-written tests. In bold, the best results for each metric. Only the test smells with non-zero values are shown in the table.\footnotemark}

\centering
\resizebox{1\textwidth}{!}{
\begin{tabular}{llrrrrrrrrrrrr}
\thead{model} & \thead{prompt\\name} & \thead{branch\\coverage\%} & \thead{line\\coverage\%} & \thead{method\\coverage\%} & \thead{mutation\\score\%} & \thead{AR} & \thead{EH} & \thead{MG} & \thead{EA} & \thead{LT} & \thead{UT} & \thead{RO} & \thead{MNT} \Tstrut\Bstrut\\ \hline
\multirow{2}{*}{gpt-4o-mini}&\multicolumn{1}{l}{zero-shot} & 41.9\%$\downarrow$ & 64.8\%$\downarrow$ & 77.2\%$\uparrow$ & 44.5\%$\uparrow$ & 1.28&0.79 & 0.19 & 2.11 & 4.52 & 0.33 & 0.11 & 7.05\Tstrut\Bstrut\\\cline{2-14}

& \multicolumn{1}{l}{few-shot} & 62.1\%$\uparrow$ & 71.3\%$\uparrow$ & 81.1\%$\uparrow$ & 61.0\%$\uparrow$ & 1.01 & 0.36 & 0.08 & 1.54 & 3.18 & 0.26 & 0.08 & 6.60 \Tstrut\Bstrut\\\hline

\multirow{2}{*}{gemini-1.5-pro} & \multicolumn{1}{l}{zero-shot} & 22.6\%$\downarrow$ & 55.4\%$\downarrow$ & 62.8\%$\downarrow$ &30.1\%$\downarrow$ & 1.67 & 0.42 &0.09 & 2.85 & 3.71 & 0.29 & 0.10 & 5.50 \Tstrut\Bstrut\\\cline{2-14}

& \multicolumn{1}{l}{few-shot} & 44.6\%$\downarrow$ & \textbf{89.8\%$\uparrow$} & \textbf{92.9\%$\uparrow$} & 72.0\%$\uparrow$ & 1.96 & 0.41 &\textbf{0.03} & 2.92 & 3.25 & \textbf{0.03} & 0.19& 5.15 \Tstrut\Bstrut\\\hline

\multirow{2}{*}{llama3.1:70b}   & \multicolumn{1}{l}{zero-shot} & 71.2\%$\uparrow$ & 80.3\%$\uparrow$ & 84.9\%$\uparrow$ & 32.0\%$\downarrow$ & 1.77 & \textbf{0.27} &0.07 & \textbf{0.51} & \textbf{1.58} & 0.49 & 0.12 & 6.20\Tstrut\Bstrut\\\cline{2-14}

& \multicolumn{1}{l}{few-shot} & \textbf{79.8\%$\uparrow$} & 85.6\%$\uparrow$ & 90.3\%$\uparrow$ & \textbf{89.2\%$\uparrow$} & \textbf{0.55} & 0.30 &0.05 & 1.31 & 1.72 & 0.10 & \textbf{0.03} & \textbf{3.85} \Tstrut\Bstrut\\\hline

\multirow{1}{*}{human}   & \multicolumn{1}{l}{-} & 48.7\% & 73.2\% & 74.0\% & 40.4\% & 2.18 & 0.69 &0.10 & 1.68 & 3.05 & 0.58 & 0.08 &4.22\Tstrut
\end{tabular}
}
\label{tab:result}
\end{table*}

\section{Evaluation and Results}
\label{sec:Discussion}
\subsection{\textbf{RQ1}: How do the chosen LLMs and prompt types perform in the test case generation task?}
We analyze each LLM/prompt combination under the evaluation protocol in Sec.~\ref{sec:eval-protocol}, i.e., full-set averaging with non-compiling generations contributing zeros. As Table~\ref{tab:result} shows, model and prompt choices materially influence all quality metrics.\footnotetext{All values are full-set averages with build failures contributing $0$. Arrows indicate whether a model--prompt configuration performs better ($\uparrow$) or worse ($\downarrow$) than the human baseline for that metric. Bold indicates the best score across all configurations.}

\subsubsection{Coverage and mutation}
The highest mutation score is achieved by \texttt{llama3.1:70b} with few-shot prompting (89.2\%), which also leads to branch coverage (79.8\%). 
The highest line and method coverages are obtained by \texttt{gemini-1.5-pro} with few-shot prompting (89.8\% and 92.9\%, respectively). 
\texttt{gpt-4o-mini} shows balanced results, with few-shot outperforming its zero-shot configuration across metrics.

\subsubsection{Test Smells}
Test smells are indicators of poor test code quality and maintainability. Table~\ref{tab:result} reports the most relevant smells. In general, LLM-generated tests are comparable to human-written ones for Assertion Roulette (AR), Mystery Guest (MG), and Resource Optimism (RO). However, they exhibit better Exception Handling (EH), particularly with \texttt{llama3.1:70b}. This model also shows fewer instances of Eager Test (EA) and Lazy Test (LT) in its zero-shot configuration. The Unknown Test (UT) smell is rare across all LLM-generated tests. Conversely, LLMs tend to use fewer hard-coded values (Magic Number Test - MNT), especially with few-shot prompts, with the exception of \texttt{gpt-4o-mini} (zero-shot).

\subsubsection{Prompt Types Comparison}
The choice between zero-shot and few-shot prompting has remarkable effects: few-shot prompts improved performance in most models compared to human-written tests, particularly in line and method coverage for \texttt{llama3.1:70b} and \texttt{gemini-1.5-pro}. This effect underscores that exposure to example inputs improves prompt generation for class-level testing tasks.
\begin{tcolorbox}[colback=gray!5!white,left=0pt,colframe=gray!40!black,title=RQ1 Findings]
\begin{itemize}
    \item \textbf{Coverage and defect rate:} \texttt{llama3.1:70b} (few-shot) with 89.2\% mutation score and 79.8\% branch coverage.
    \item \textbf{Effect of context:} The \texttt{gemini-1.5-pro} model prompt performed well in line and method coverage when using few-shot prompts, highlighting the importance of providing contextual examples.
    \item \textbf{LLM vs human:} LLM-generated tests matched or exceeded human-written tests on some quality metrics, when considering only the subset of tests that compile.
\end{itemize}
\end{tcolorbox}

\begin{table}[ht]
\caption{Percentage of build success on the full set per experiment}
\centering
\begin{tabular}{lcrrr}
\thead{model}                   & \thead{prompt\\name}           & \thead{Build\%}\Tstrut\Bstrut\\\hline
\multirow{2}{*}{gpt-4o-mini}    & \multicolumn{1}{l}{zero-shot} & 28.6\%\Tstrut\Bstrut\\\cline{2-3}
                                & \multicolumn{1}{l}{few-shot}  & 25.3\% \Tstrut\Bstrut\\\hline
\multirow{2}{*}{\textbf{gemini-1.5-pro}} & \multicolumn{1}{l}{zero-shot} & 18.6\%\Tstrut\Bstrut\\\cline{2-3}
                                & \multicolumn{1}{l}{\textbf{few-shot}}  & \textbf{36.0\%} \Tstrut\Bstrut\\\hline
\multirow{2}{*}{llama3.1:70b}   & \multicolumn{1}{l}{zero-shot} & 9.8\%\Tstrut\Bstrut\\\cline{2-3}
                                & \multicolumn{1}{l}{few-shot}  & 7.1\%\Tstrut\Bstrut\\\hline
\multirow{1}{*}{human}          & \multicolumn{1}{l}{-}         & \multicolumn{1}{r}{100\%}\Tstrut
\end{tabular}
\label{tab:compilation}
\end{table}

\subsection{\textbf{RQ2}: How frequently do compilation errors occur, and how do they impact the overall compilation success rate?}
After the initial experimental run, we observed relatively low compilation success rates across all tested LLMs, limiting the effective coverage analysis to successfully compiled classes. Therefore, the coverage metrics reported reflect only this subset of valid tests (see Table \ref{tab:compilation}).
To investigate this issue, we systematically collected and analyzed errors that prevented a successful compilation during our experiments. Two developers worked to categorize the errors. Each developer received a list of errors along with the corresponding generated code that triggered these errors. They independently labeled the errors without consulting each other. Upon completing their independent labeling, they joined to compare and discuss the results.
A reconciliation process was conducted to resolve discrepancies in their labels. If there are any mismatches, the two developers discuss the differences to reach a consensus. In cases where they could not agree, a third developer acted as an arbitrator to make the final decision. 
The analysis of the errors revealed several recurring patterns that impacted the compilation success rate. These errors can be grouped into three main categories:

\subsubsection{Symbol and Reference Issues}
\paragraph{Cannot Find Symbol} This was the most frequent error, occurring in 42.50\% of cases. It typically involves references to variables, methods, or classes that are not defined or incorrectly named. This highlights the challenges the LLM faced in accurately recalling or generating the necessary components of the code.
\paragraph{Missing or Incorrect Imports} Errors due to missing import statements or incorrect library versions account for 19.55\% of the total errors. This issue often arose from a mismatch between the generated code and the actual project dependencies, highlighting limitations in the models' understanding of the project environment.
\subsubsection{Code Structure and Consistency Issues}
\paragraph{Override/Implementation Issues} Errors related to incorrect overrides or failed implementation of interfaces contribute to 13.87\% of the errors. These errors often occur when the generated methods do not match the expected signatures or fail to fulfill the requirements of implemented interfaces.
\paragraph{Visibility/Access Issues} These errors, making up 7.69\%, are due to incorrect access modifiers or attempts to access private or protected members from outside their intended scope. This indicates that the model sometimes struggled with correctly applying Java's access control rules.
\paragraph{Incorrect Data Types} Errors also arise from mismatched data types, accounting for 6.31\% of the cases. This typically happens when the generated code attempts to pass arguments of one type to methods expecting a different type, suggesting challenges in maintaining type consistency across the generated test suite.
\paragraph{Instantiation Issues} Errors involving incorrect object instantiation account for 3.66\% of the total. These often come up when the generated code uses the wrong constructor or attempts to instantiate abstract classes.

\subsubsection{Syntax and Specific Rule Violations}
\paragraph{Syntax Errors} These include incorrect method signatures, missing semicolons, and other syntactic issues, making up 5.17\% of the errors. Such errors suggest that, while LLMs can generate logically coherent code, they occasionally fail to adhere to the strict syntactical rules of Java.
\paragraph{Final Variable Issues} Errors related to improper use of final variables represented 1.26\% of the cases. These typically involve attempts to reassign values to final variables, reflecting misunderstandings in handling Java's immutability constraints.
The \texttt{gemini-1.5-pro} model stands out for achieving the highest compilation rate among the models tested, reaching 36.0\% in the few-shot configuration. This outcome highlights the need to increase the compilation rate further, which we address in RQ3.

\begin{tcolorbox}[colback=gray!5!white,left=0pt,colframe=gray!40!black, title=RQ2 Findings]
\begin{itemize}
\item The most common issues affecting the compilation success rate are related to missing symbols and incorrect references (62.05\%).
\item Code structure problems, including overrides, access control, and data types, contributed 31.53\%.
\item Syntax and specific rule violations, such as improper use of final variables, accounted for 6.43\%.
\end{itemize}
\end{tcolorbox}

\subsection{\textbf{RQ3}: Is there a strategy to increase the compilation success rate?}
After analyzing the distribution of errors identified in the generated test suites, we observed that a significant portion of compilation failures were attributed to missing symbols or incorrect references, categorized under symbols and references issues. To address the compilation rate issues, we implemented an enhanced prompting strategy, explicitly providing the full path of classes under test within prompts. This refinement significantly increased compilation success rates by enabling LLMs to generate more contextually accurate import statements and symbol references.
The enhanced strategy involves expanding the set of configuration parameters provided to the LLM by adding a new parameter, called ``\texttt{class\_under\_test\_path}”. This parameter explicitly specifies the path to the class under test within the project, providing a clearer reference to ensure that the model has access to accurate and correct imports and symbol definitions related to the class under test. In addition, the prompts were updated to include this parameter, which provided a more grounded understanding of the model with respect to the dependencies and structure of the class under test. A compact before$\rightarrow$after summary is reported in \Cref{tab:comparison}.










\begin{table}[t]
\caption{Compact per-model before$\rightarrow$after deltas. Values are absolute percentage-point changes.}
\centering
\small
\resizebox{1\linewidth}{!}{
\begin{tabular}{llrrrrr}
\thead{model} & \thead{prompt} & \thead{$\Delta$ Build (pp)} & \thead{$\Delta$ Branch} & \thead{$\Delta$ Line} & \thead{$\Delta$ Method} & \thead{$\Delta$ Mutation}\\
\hline
\multirow{2}{*}{gpt-4o-mini}    & zero-shot  & +28.3 & +3.9 & +3.7 & +2.8 & +4.1 \\
& few-shot   & +26.8 & +3.1 & +3.4 & +2.5 & +3.2 \\\hline
\multirow{2}{*}{gemini-1.5-pro} & zero-shot  & +6.2  & +1.3 & +1.9 & +1.5 & +1.7 \\
& few-shot   & +28.6 & +2.0 & +2.2 & +1.8 & +2.6 \\\hline
\multirow{2}{*}{llama3.1:70b}   & zero-shot  & +14.3 & +1.8 & +1.6 & +1.2 & +2.1 \\
& few-shot   & +13.8 & +1.9 & +2.0 & +1.6 & +2.7 \\
\end{tabular}
}
\label{tab:comparison}
\end{table}

These improvements demonstrate that providing more precise details during the test generation process can significantly increase the chances of generating compilable code. The focus on explicit class under test paths allowed the LLMs to better handle project-specific imports and references, thereby addressing one of the major limitations observed in our initial experiments.

\begin{tcolorbox}[colback=gray!5!white,left=0pt,colframe=gray!40!black, title=RQ3 Findings]
\begin{itemize}
\item The enhanced strategy consistently raises build rates for all models and yields small but systematic increases in coverage and mutation score.
\item Explicitly specifying the class under test path provides a more accurate reference, reducing errors related to missing symbols and incorrect references.
\item Despite these improvements, additional strategies (e.g., automated post-generation repair or advanced context-aware prompting) are still needed to approach higher compilation success rates and functional correctness.
\end{itemize}
\end{tcolorbox}

\section{Limitations}
\label{sec:limitations}
Although \textsc{AgoneTest} presents an innovative framework for automating the generation and evaluation of unit test suites using LLMs, we acknowledged some limitations of its current first implementation and experimental results. \paragraph{Compilation Success Rate} The low compilation success rates observed represent a limitation inherent in the LLM-based code generation process rather than the \textsc{AgoneTest} framework itself. These limitations stem from the LLMs' challenges in understanding complex contextual dependencies. Future work could investigate new techniques to improve compilation rates, potentially by extending \textsc{AgoneTest} to incorporate automated code-repair strategies. \paragraph{Dataset and Generalization} For our evaluation, we relied on a newly created \textsc{Classes2Test} dataset which includes Java projects. This makes our findings not immediately generalizable to different programming languages. Moreover, the repositories included in \textsc{Classes2Test} were selected based on their ability to compile without errors, potentially introducing a bias towards well-structured codebases. \paragraph{Quality Metrics} Coverage, mutation score, and test smells are proxies of test effectiveness; they do not fully capture functional correctness. \paragraph{Data Leakage} Data leakage is a concern when evaluating LLMs. We mitigate this by using a recently created dataset with projects updated after the training cutoff dates of the models used. However, this could become a more significant problem as models evolve. \paragraph{Prompting} Our goal is an automated evaluation framework for LLM-based unit test generation; current experiments primarily illustrate how \textsc{AgoneTest} functions. The framework allows selecting LLMs and specifying prompts, but we explored only two prompting strategies, limiting full insight into model capabilities. Additionally, manual exemplar selection in one-shot prompts introduces subjectivity, raising concerns about prompt sensitivity and potential overfitting rather than genuine generalization. \paragraph{Randomness/repeated runs} Results are from a single run with temperature=0; repeated-run variance analysis is left to future work and is supported by \textsc{AgoneTest}’s temperature setting.

\section{Conclusion and Future Work}
\label{sec:conclusion}
We introduced \textsc{AgoneTest}, an automated framework for evaluating unit test suites generated by LLMs on real-world Java projects. The main contribution of \textsc{AgoneTest} lies in its ability to provide a reproducible and extensible evaluation pipeline, integrating project setup, execution of generated tests, and quality metrics computation.
To support this evaluation, we introduced the \textsc{Classes2Test} dataset, which enables systematic benchmarking at the class level and complements existing resources such as \textsc{Methods2Test}.
By focusing on evaluation rather than generation, \textsc{AgoneTest} offers the community a tool to compare LLMs, prompting strategies, and future improvements in test automation. As illustrated by our motivating use case, it transforms the ad hoc, error-prone process of manual evaluation into a systematic and reproducible workflow for developers and researchers.
Our findings indicate that LLM-generated tests achieve comparable or superior code coverage and defect detection compared to human-written tests, particularly when context-aware prompting is used. This highlights the potential of LLMs in automating software testing, despite existing limitations.
Future work will enhance \textsc{AgoneTest} by supporting more languages and boosting compilation rates. We also plan to expand the experimental scope with a wider variety of LLMs and advanced prompting techniques. Such enhancements will further empower developers, as described in our use case, to not only benchmark models but also to select and refine the optimal test generation strategy for their specific projects, thereby improving overall software quality.

\section*{Acknowledgment}
This work was funded by: the European Union -- NextGenerationEU under the \textit{P+ARTS -- Partnership for Artistic Research in Technology and Sustainability} (NRRP -- M4C1, Investment 3.4, INTAFAM00037; CUP: G43C24000640006), \textit{PE9 GRINS -- Growing Resilient, INclusive and Sustainable} project ``VIRAL Data Engine -- Virtualization and Intelligence Resource for Advanced Learning'' (NRRP -- M4C2, Investment 1.3, PE00000018; CUP: J33C22002910001), PRIN 2022 - ERC PE6 ``TRex-SE: Trustworthy
Recommenders for Software Engineers'' (2022LKJWHC\_03 - CUP: D53D23008730006) and PRIN 2022 - ERC SH5, PE6 ``The Words of Peace and Pacifism. French Literature in the Inter-war period by exploiting Distributional Semantic Analysis'' (P20228AMFB - CUP: D53D23019570001)

\bibliographystyle{IEEEtran}
\bibliography{bibliography}
\end{document}